# Design and Implementation of Modified Fuzzy based CPU Scheduling Algorithm

Rajani Kumari
Jagannath University
Chaksu, Jaipur,
Rajasthan, India--303901

Vivek Kumar Sharma, Ph.D
Jagannath University
Chaksu, Jaipur,
Rajasthan, India--303901

Sandeep Kumar
Jagannath University
Chaksu, Jaipur,
Rajasthan, India--303901

## ABSTRACT

CPU Scheduling is the base of multiprogramming. Scheduling is a process which decides order of task from a set of multiple tasks that are ready to execute. There are number of CPU scheduling algorithms available, but it is very difficult task to decide which one is better. This paper discusses the design and implementation of modified fuzzy based CPU scheduling algorithm. This paper present a new set of fuzzy rules. It demonstrates that scheduling done with new priority improves average waiting time and average turnaround time.

## 1. INTRODUCTION

Scheduling is a method by which thread or process are granted access to computer resources. Operating system is responsible for scheduling a process. Scheduling becomes necessary and important task when we have more than one ready to execute job. Scheduling optimize use of resources. Criteria for choosing best scheduling algorithm vary according nature or preferences of task. Five basic features to decide best scheduling algorithm are as follow [1, 2]:

- Utilization of CPU time
- Throughput
- Turnaround Time
- Waiting Time
- Response time

This paper considers only average waiting time and average turnaround time for the purpose of performance comparison. The organization of rest of paper is as follow: In section 2, we introduce three scheduling techniques named SJF, Priority scheduling and Fuzzy based priority scheduling. Next section proposes a modified fuzzy based scheduling algorithm and set of fuzzy rules. Section 4 gives results based on some case studies. Here one case study taken from paper presented by Kadhim, Shatha J., and Kasim M. Al-Aubidy in [1]. Finally conclude the result in section 5.

## 2. SCHEDULING TECHNIQUES

Scheduling algorithms [3, 4] can be classified in different categories as per their applications. Like:

**Preemptive/ Non-preemptive Techniques** – In case of preemptive scheduling algorithm, scheduler can stop/postpone execution of any task. But in case of non-preemptive algorithms it is not possible till job completion or running job voluntarily give up resources [2, 3].

**Static /Dynamic priority scheduling** – Static priority do not change, but dynamic priority can be modified.

This paper considers three scheduling algorithms Shortest Job First, Priority Scheduling and Fuzzy based CPU scheduling.

**Shortest Job First (SJF) Scheduling Algorithm:** This algorithm schedules jobs according to their execution time, job with lowest execution time scheduled first and job with largest execution time scheduled last. This is a non-preemptive algorithm. Its variant known as shortest remaining time is preemptive. This algorithm reduces average waiting time. Shortest job first can be effectively used in case of interactive processes that usually follow a pattern of alternating between looking ahead to a command and execution of it. If the execution time of a method is considered a separate "job", past behavior will indicate that method to run next, supported based on estimate of its time period [2, 3, 5].

**Priority Scheduling Algorithm:** The Operating System decides priority to each method, and therefore the scheduler arranges the processes within the ready queue so as of their priority. Processes with low priority may be interrupted when incoming process has high priority. There is a number of variant of priority scheduling; SJF is also a special case of priority scheduling. This algorithm sometimes face problem of infinite blocking and starvation, this problem can be removed by aging technique [2, 3].

**Fuzzy based scheduling:** Fuzzy based scheduling algorithm proposed by Kadhim, Shatha J., and Kasim M. Al-Aubidy in [1]. This algorithm takes input both job priority and execution time and decides new priority using some fuzzy rules. This algorithm uses some linguistic variables for priority given by system, new computed priority and execution time. Performance of this algorithm compared with SJF and classical priority scheduling methods [1].

## 3. FUZZY LOGIC

Fuzzy logic is a style of multi-valued logic. It deals with approximation rather than exactness. In contrast to classical sets (Classical set takes true or false values) fuzzy logic variables (also known as linguistic variable) can have a truth value that ranges in interval between 0 and 1. Fuzzy logic has been prolonged to grasp the concept of fractional truth, where the truth value may range between completely true and completely false. Moreover, when linguistic variables are practiced, these degrees may be determined by specific methods [6].

In recent years the application of fuzzy systems in real world issues is increasing due to the actual fact that fuzzy systems will affect linguistic knowledge beside the numerical knowledge. Now a day an outsized range of researchers are shifting towards fuzzy logic. Fuzzy logic has shown itself to be a strong design and analysis methodology in control theory, enabling the implementation of advanced knowledge-based control methods for complicated dynamic systems like those rising applications for systems and artificial biology. TA Hiwarkar et al. list a wide range of applications of fuzzy logic in [6]. K.A. Verma in [7] discussed type-1 fuzzy system and origin of type-2 fuzzy systems and theories and application fields of engineering, finance and medical domains. Xia Feng et al. [8] designed scheduler for embedded control system using fuzzy logic based feedback. A priority scheduler has been developed for mobile ad hoc networks by C. Gomathy et al. in [9]. Rajani Kumari et al. discussed air conditioning system with fuzzy logic [10]. A. Saleh proposed a grid-scheduling algorithm on fuzzy matchmaking approach [11].

Fuzzy logic based algorithms are getting more popularity now a day as Reza salami et al. developed an efficient task scheduling algorithm for computational grids using NSGA II with fuzzy variance based crossover [12]. In this algorithm two functions are defined to generate two inputs for fuzzy based system. Variance of costs and presence of resource in scheduling are used to specify probability of crossover. It provides better solution in less number of iterations. H. Chuan et al. discussed fuzzy job shop scheduling problem based in interval number theory [13]. S. Mandloi et al. gives new idea based on ant colony optimization with genetic parameter selection for job scheduling in computational grid [14]. Fuzzy based scheduling algorithms implemented in cloud also. C. Yi Chun et al. proposed a fuzzy based dynamic load decision making scheme in cloud computing [15].

## 4. PROPOSED ALGORITHM

This scheduling deal with some fuzzy rules and these rules are based on assigned priority and execution time. This work is proposed to compute the New Priority (NP) for all tasks using pre priority and execution time with the help of Mamdani type inference [16].

This paper use suitable linguistic variables as input and output for compute a crisp value for new priority. Pre Priority (PP) measured as Very Low, Low, Medium, High and Very High. Execution Time (ET) measured as Very Small, Small, Medium, Long and Very Long. New Priority (NP) measured as Very Low, Low, Medium, High and Very High. The proposed scheduling is a collection of linguistic fuzzy rules which describe the relationship between defined input variables (PP and ET) and output (NP).

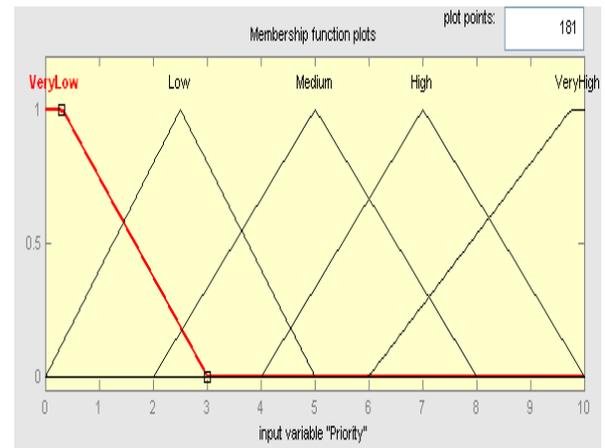

**Fig. 1: Membership Function for Priority**

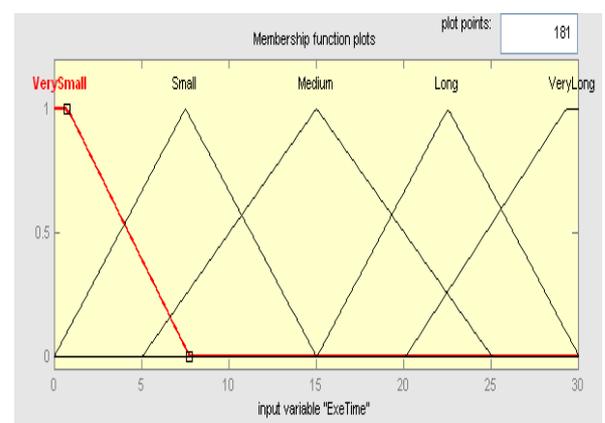

**Fig. 2: Membership Function for Execution Time**

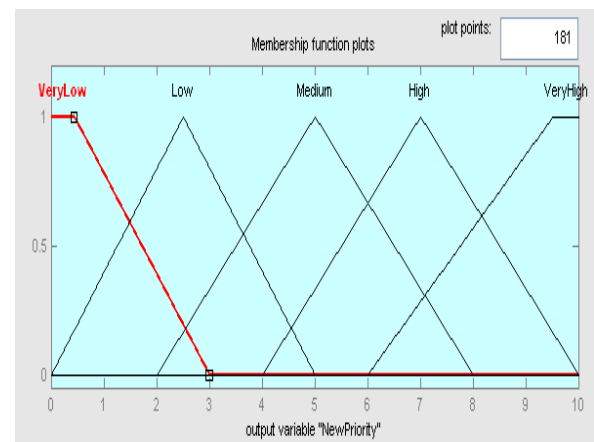

**Fig. 3: Membership Function for New Priority**

Table contains 25 rules which are based on IF THEN statement such as: -

**IF PP is High and ET is Small THEN NP is High**

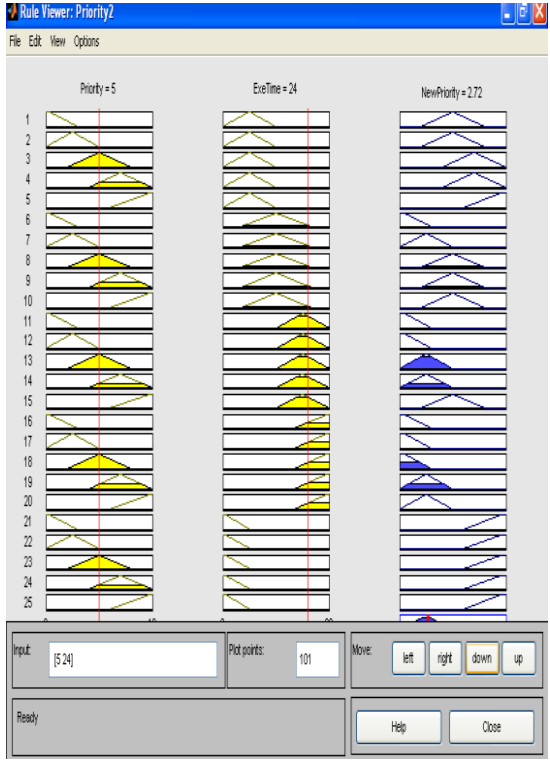

**Fig. 4: Computation of New Priority**

**Table 1. Fuzzy rules for proposed design**

| PRIORITY | EXECUTION TIME | NEW PRIORITY |
|---|---|---|
| Very Low | Very Small | Very High |
| Low | Very Small | Very High |
| Medium | Very Small | Very High |
| High | Very Small | Very High |
| Very High | Very Small | Very High |
| Very Low | Small | Medium |
| Low | Small | Medium |
| Medium | Small | High |
| High | Small | High |
| Very High | Small | Very High |
| Very Low | Medium | Very Low |
| Low | Medium | Low |
| Medium | Medium | Medium |
| High | Medium | Medium |
| Very High | Medium | Medium |
| Very Low | Long | Very Low |
| Low | Long | Very Low |
| Medium | Long | Low |
| High | Long | Low |
| Very High | Long | Low |
| Very Low | Very Long | Very Low |
| Low | Very Long | Very Low |
| Medium | Very Long | Very Low |
| High | Very Long | Low |
| Very High | Very Long | Low |

These rules compute the crisp value using centroid Defuzzification method of Mamdani inference in MATLAB that represents the NP of each task.

For example: - If Priority is 5 and Execution Time is 20 Then New Priority is 3.81

After assigning new priority, execution is start according to assigned priority and arrival time. Firstly find out the entire tasks which have minimum arrival time (0 sec) and execute the task which has higher priority from them till next task will enter in queue. Whenever a new task will enter again check the priority of entered task and execute the higher priority task till next tasks will enter and so on. After entering all tasks in queue execute remaining task as per higher to lower priority.

**ALGORITHM 1: Fuzzy based CPU scheduling**

1.) Initialize $n$ processes with their Execution Time, Priority and Arrival Time.
2.) Compute New Priority ($NP$) using fuzzy rules with the help of Execution Time and Priority.
3.) Apply sorting method on arrival time to arrange in ascending order.

```
for (i=1; i<n; i++)
{
for (j=1; j<n-i; j++)
if (AT_nj>AT_nj+1)
swap
```

4.) Compute the frequency of same arrival time and store in x variable.
```
z=1, AT = AT_ni,    x=1
for (i=z; a<=n; i++)
        if (AT_ni == AT_ni+1)
        AT = AT_ni, x=x+1, z=z+1
        Else
        AT1 =AT _ni+1, z=z+1
Break
```
5.) Compare New Priority of all the process which is having arrival time 0 and find Highest Priority (HP).
```
HP= NP_ni
for (i=1; i<=x; i++) x  is  frequency  of
arrival time 0.
if (NP_ni < NP_ni+1)
        HP = NP_ni+1
        Execute the process which has
        Highest Priority.
```
6.) Execute this process until new process will not enter in queue.
```
for (i=1; i<AT1-AT; i++)
Execute P
```
7.) Whenever new process is enter, break execution of this process and then again find highest priority among all entered processes.
8.) After entering all processes in queue apply sorting method on New Priority to arrange in ascending order.
9.) Execute all processes as per assigned priority.

It is very clear that the proposed fuzzy based algorithm obtain better result rather than another scheduling algorithm. It reduces the average waiting time and average turnaround time.

# 5. RESULT AND PERFORMANCE OF MODIFIED FUZZY BASED SCHEDULING

To evaluate the performance and applicability of fuzzy based scheduling algorithm, we have taken different case studies and compare some scheduling algorithms such as SFJ, Priority algorithm, Fuzzy based CPU scheduling algorithm and proposed fuzzy based scheduling algorithm. Case study 1 taken from [1] and case study 2 is an independent example. Table 2 contains a case study with five different processes with different execution time and priority. It is assumed that all process available at time 0.

**Table 2: Case Study-1(Without Arrival Time)**

| Processes | P1 | P2 | P3 | P4 | P5 |
|---|---|---|---|---|---|
| Execution time | 3 | 24 | 6 | 9 | 8 |
| Priority | 6 | 5 | 1 | 4 | 2 |
| Fuzzy based Priority | 5.961 | 4.407 | 4.891 | 5.081 | 4.967 |
| New Priority | 7.66 | 2.72 | 5.41 | 5.31 | 4.22 |

1) Gantt chart for Priority Scheduling:

| P1 | P2 | P4 | P5 | P3 |
|---|---|---|---|---|
| 0 | 3 | 27 | 36 | 44 | 50 |

Average Waiting Time: 22
Average Turn Around Time: 32

2) Gantt chart for SJF Scheduling:

| P1 | P3 | P5 | P4 | P2 |
|---|---|---|---|---|
| 0 | 3 | 9 | 17 | 26 | 50 |

Average Waiting Time: 11.0
Average Turn Around Time: 21.0

3) Gantt chart for Priority using Fuzzy based CPU Scheduling:

| P1 | P4 | P5 | P3 | P2 |
|---|---|---|---|---|
| 0 | 3 | 12 | 20 | 26 | 50 |

Average Waiting Time: 12.2
Average Turn Around Time: 22.2

4) Gantt chart for New Priority Scheduling:

| P1 | P3 | P4 | P5 | P2 |
|---|---|---|---|---|
| 0 | 3 | 9 | 18 | 26 | 50 |

Average Waiting Time: 11.2
Average Turn Around Time: 21.2

Comparison of proposed algorithm with SJF, Priority scheduling and fuzzy based priority scheduling based on case study-1(Without Arrival Time) outlined in Fig. 5. It is clear from results that new priority algorithm outperform almost all existing algorithms.

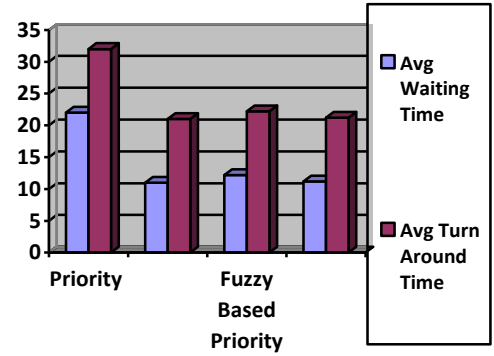

**Fig. 5: Result of Case study-1(Without Arrival time)**

Table 3 again contains a case study with five different processes with different execution time and priority. Now each process assigned arrival time.

**Table 3: Case Study-1(With Arrival Time)**

| Processes | P1 | P2 | P3 | P4 | P5 |
|---|---|---|---|---|---|
| Execution time | 3 | 24 | 6 | 9 | 8 |
| Arrival Time | 2 | 1 | 2 | 1 | 0 |
| Priority | 6 | 5 | 1 | 4 | 2 |
| Fuzzy based Priority | 5.961 | 4.407 | 4.891 | 5.081 | 4.967 |
| New Priority | 7.66 | 2.72 | 5.41 | 5.31 | 4.22 |

1) Gantt chart for Priority Scheduling:

| P5 | P1 | P2 | P4 | P3 |
|---|---|---|---|---|
| 0 | 8 | 11 | 35 | 44 | 50 |

Average Waiting Time: 18.4
Average Turn Around Time: 28.4

2) Gantt chart for SJF Scheduling:

| P5 | P1 | P3 | P4 | P2 |
|---|---|---|---|---|
| 0 | 8 | 11 | 17 | 26 | 50 |

Average Waiting Time: 11.2
Average Turn Around Time: 21.2

3) Gantt chart for Priority using Fuzzy based CPU Scheduling:

| P5 | P1 | P4 | P3 | P2 |
|---|---|---|---|---|
| 0 | 8 | 11 | 20 | 26 | 50 |

Average Waiting Time: 11.8
Average Turn Around Time: 21.8

4) Gantt chart for New Priority Scheduling:

| P5 | P4 | P1 | P3 | P4 | P5 | P2 |
|---|---|---|---|---|---|---|
| 0 | 1 | 2 | 5 | 11 | 19 | 26 | 50 |

Average Waiting Time: 11.4
Average Turn Around Time: 21.00

Comparison of proposed algorithm with SJF, Priority scheduling and fuzzy based priority scheduling based on case study-1(With Arrival Time) shown in Fig. 6. It demonstrates that proposed algorithm outperform mostly existing algorithms.

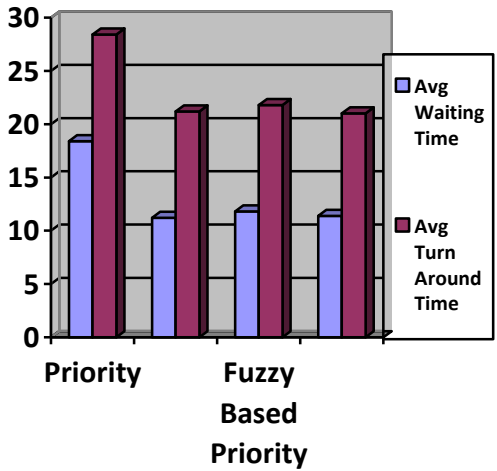

**Fig. 6: Result of Case study-1(With Arrival time)**

Table 4 contains an independent case study with eight different processes with different execution time and priority. Each process assigned arrival time.

**Table 4: Case Study-2**

| Processes | P1 | P2 | P3 | P4 | P5 | P6 | P7 | P8 |
|---|---|---|---|---|---|---|---|---|
| Execution time | 18 | 5 | 7 | 2 | 9 | 20 | 12 | 15 |
| Arrival Time | 4 | 0 | 2 | 2 | 0 | 2 | 0 | 4 |
| Priority | 5 | 6 | 3 | 7 | 4 | 8 | 2 | 9 |
| New Priority | 4.2 | 7.23 | 5.24 | 7.96 | 5.31 | 3.84 | 3.49 | 5 |

1)   Gantt chart for Priority Scheduling:

| P2 | P6 | P8 | P6 | P4 | P2 | P1 | P5 | P3 | P7 |
|---|---|---|---|---|---|---|---|---|---|

Average Waiting Time: 41
Average Turn Around Time: 52

2)   Gantt chart for SJF Scheduling:

| P2 | P4 | P3 | P5 | P7 | P8 | P1 | P6 |
|---|---|---|---|---|---|---|---|

Average Waiting Time: 23.5
Average Turn Around Time: 34.5

3)   Gantt chart for New Priority Scheduling:

| P2 | P4 | P2 | P5 | P3 | P8 | P1 | P6 | P7 |
|---|---|---|---|---|---|---|---|---|

Average Waiting Time: 25.5
Average Turn Around Time: 36.75

Comparative analysis of proposed algorithm with SJF, Priority scheduling and fuzzy based priority scheduling based on case study-2 shown in Fig. 7. It is outlined that proposed algorithm outperform priority scheduling and a little bit less than SJF algorithms.

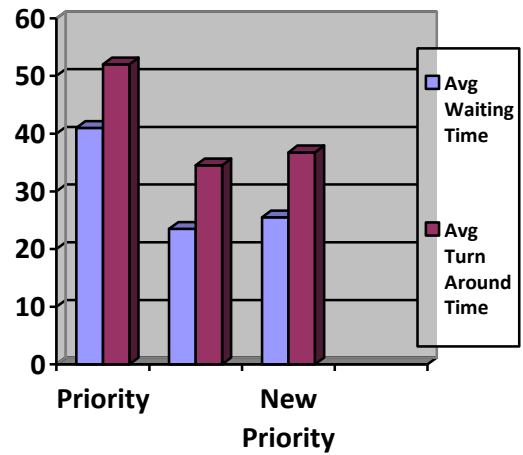

**Fig. 7: Result of Case study-2**

# 6. CONCLUSIONS

The proposed fuzzy based scheduling algorithm is an efficient scheduling algorithm that is obtained batter result rather than other algorithm. Figure 5 and 6 shows the comparison between SJF, Priority scheduling algorithm, Fuzzy based CPU scheduling algorithm and proposed new fuzzy based scheduling algorithm. The average waiting time and average turnaround time of proposed fuzzy based algorithm is much better than the Priority algorithm, Fuzzy based CPU scheduling algorithm and closer to obtain by SJF algorithm, but SJF algorithm doesn't deal with task priority. Figure 7 shows the comparison between SJF, Fuzzy based CPU scheduling algorithm and proposed new fuzzy based scheduling algorithm. Results prove that algorithm proposed in this paper is much better than existing algorithms.